\documentclass[12pt,a4paper]{article}
\usepackage{amsmath,amssymb,type1cm,booktabs,bm,braket}
\usepackage[dvips]{graphicx}
\usepackage{amsmath,amssymb}
\usepackage{comment}	
\usepackage{mathrsfs}
\usepackage{bm}
\usepackage{ascmac}
\usepackage{enumerate}
\usepackage{cancel}
\graphicspath{{./figures/}}

\allowdisplaybreaks[1]
\numberwithin{equation}{section}

\DeclareMathOperator{\direct}{direct}

\DeclareMathOperator{\mev}{MeV}
\DeclareMathOperator{\gev}{GeV}
\DeclareMathOperator{\tev}{TeV}

\def\Lim{\qopname\relax\@empty{lim}\limits}
\let\lim = \Lim

\def\sup{\mathop{\operator@font sup}\limits}
\def\inf{\mathop{\operator@font inf}\limits}
\def\max{\mathop{\operator@font max}\limits}
\def\min{\mathop{\operator@font min}\limits}
\def\prod{\mathop{\mathchoice{\textstyle\prod@}{\textstyle\prod@}{%
      \scriptstyle\prod@}{\scriptscriptstyle\prod@}}\limits}
\def\coprod{\mathop{\mathchoice{\textstyle\coprod@}{\textstyle\coprod@}{%
      \scriptstyle\coprod@}{\scriptscriptstyle\coprod@}}\limits}
\def\bigcap{\mathop{\mathchoice{\textstyle\bigcap@}{\textstyle\bigcap@}{%
      \scriptstyle\bigcap@}{\scriptscriptstyle\bigcap@}}\limits}
\def\bigcup{\mathop{\mathchoice{\textstyle\bigcup@}{\textstyle\bigcup@}{%
      \scriptstyle\bigcup@}{\scriptscriptstyle\bigcup@}}\limits}
\def\Int{\displaystyle\intop\ilimits@}
\def\bigoplus{\mathop{\mathchoice{\textstyle\bigoplus@}{%
      \textstyle\bigoplus@}{\scriptstyle\bigoplus@}{%
      \scriptscriptstyle\bigoplus@}}\limits}
\def\slashchar#1{\setbox0=\hbox{$#1$}	
\dimen0=\wd0				
\setbox1=\hbox{/} \dimen1=\wd1		
\ifdim\dimen0>\dimen1			
\rlap{\hbox to \dimen0{\hfil/\hfil}}	
#1					
\else 					
\rlap{\hbox to \dimen1{\hfil$#1$\hfil}}	
/					
\fi}
\begin{document}
\begin{titlepage}

 \begin{flushright}
 \end{flushright}

 \vspace{1ex}

 \begin{center}

  {\LARGE\bf Averaged Number of the Lightest Supersymmetric Particles in Decay 
  of Superheavy Particle with Long Lifetime}

  \vspace{3ex}

  {\large $^a$Yasuhiro Kurata
  \footnote{e-mail: yas@eken.phys.nagoya-u.ac.jp},
  $^{ab}$Nobuhiro Maekawa
  \footnote{e-mail: maekawa@eken.phys.nagoya-u.ac.jp}}

  \vspace{4ex}
  {\it$^a$ Department of Physics, Nagoya University, Nagoya 464-8602, Japan}
  \\
  {\it$^b$ Kobayashi Maskawa Institute, Nagoya University, Nagoya 464-8602, Japan}
  \\
  \vspace{6ex}

 \end{center}

 \begin{abstract}
 We calculate the averaged number $\nu$ of the lightest supersymmetric particles (LSPs)
 in a shower from the decay of superheavy particle $X$ by generalized DGLAP
 equations. If the primary decayed particles have color charges and the virtuality
 is around $10^{13}-10^{14}$ GeV, the averaged
 number of the LSPs can become O(100).
 As the result, the upper limit of the mass of the superheavy particle, 
 whose decay can produce the observed abundance of the dark matter, can increase from
 $10^{12}$ GeV to $10^{14}$ GeV. Since the typical scale of the inflaton mass of
 the chaotic inflation is around $10^{13}$ GeV, the decay of the inflaton
 can produce the observed dark matter abundance if the reheating temperature is of order
 1 GeV.  
 Even for the standard model particles with virtuality $Q\sim 10-100$ TeV,
 the averaged number of the LSPs becomes O(0.1) for gluon, and O(0.01) for Higgs,
 which strongly constrains the scenario of non-thermal LSP production from the decay
 of moduli with 10-100 TeV mass.
\end{abstract}

\end{titlepage}
\section{Introduction}
%
Recently, the amount of the dark matter (DM) in the universe has been precisely measured 
by WMAP observation\cite{WMAP} as
\begin{equation}
\Omega_{DM}h^2=0.1120\pm0.0036.
\end{equation}
However, the origin of the DM is still unknown, and it is an important challenge to
reveal the origin of the DM.
One of the most promising candidates for the dark matter is weakly interacting massive
particle (WIMP), because the thermal production can explain the observed amount and such 
a particle is required to solve the naturalness problem on the standard model (SM) Higgs.
For example, in the minimal supersymmetric (SUSY) SM (MSSM), the lightest SUSY paricle
(LSP) becomes stable because of R-parity. And if the LSP is a neutralino, 
it becomes a candidate of the WIMP. We can check this scenario if the thermal abundance is calculated by the parameters
measured by the experiments. However, in principle, it may happen that the calculated 
thermal abundance is inconsistent with the observed value. Moreover, even in the SUSY models,
there is a possibility that the LSP has no weak interaction like the gravitino or axino.
 Therefore, it is worth considering other possibilities for the production of the LSP.

In the literature, a lot of studies on the non-thermal production of
the LSP have been studied\cite{NTDM1}-\cite{NTDM8}. The energy density of the LSP $\chi$ originated from the decay
of the superheavy field $X$ at the decay of $X$ is estimated as 
\begin{equation}
\rho_{\chi}=m_\chi n_{\chi}\sim m_\chi\nu n_X,
\end{equation}
where $m_{\chi}$, $n_{\chi}$, and $n_X$ are the mass of the LSP, the number density of the LSP, and
the number density of the superheavy field $X$. $\nu$ is an averaged number of produced LSPs
from the decay of a single superheavy field $X$. The parameter $\nu$ is important to obtain the final 
result, though it is dependent on the explicit models.
In many cases, the relation $\nu\sim 1$ is adopted.
This is because when the R-parity of the X particle is odd, 
it is obvious that the number of the produced
 sparticle is around one in the lowest order calculation, 
and when the R-parity is even, the branching ratio to a pair of sparticles becomes
the same order as the branching ratio
to the SM particles in the lowest order calculation due to the supersymmetry.
However, it has been discussed that the $\nu$ can be much smaller than 1 in the special cases.
For example, the branching ratio for the decay of the moduli with 10-100 TeV mass 
into the gaugino pair can be much smaller than
into the gauge boson pair because of chirality suppression\cite{NTDM3}. 
In the scenario, the small $\nu\sim 10^{-4}$ is required
to obtain the observed abundance of the Wino-like LSP.

In this paper, we reconsider how many LSPs are
produced by one superheavy particle decay.
Basically, the primary particle in the decay must produce a shower 
at least if the primary particle
has color charges. And if the mass of the superheavy particle is much larger than the SUSY breaking
scale, then it is naturally expected that in the shower, sparticles can be produced.
We estimate the number of the LSPs in the shower by using supersymmetric 
Dokshitzer-Gribov-Lipatov-Altarelli-Parisi(DGLAP) equations\cite{DGLAP,SarkarToldra,SHDecay}.
Of course, the number
of the LSPs in the shower is dependent on the mother MSSM particles produced by the decay and
on the mass scale of the superheavy particle. We make several tables so that our calculation can be applied
to various masses of the superheavy particles and to various decay modes. 

In section 2, we briefly remind the scenario on the non-thermal production of the LSP.
In section 3, we calculate the parameter $\nu$ numerically.
The  section 4 is for the discussions and summary.

\section{Non thermal production of LSP}
In this section, we estimate the LSP production by the decay of the superheavy particle $X$.
We assume that the decay modes are some of the MSSM particles. 
Supposing that the superheavy particle $X$ dominates the energy density of the
universe at the decay, the radiation energy density $\rho_{\text{rad}}$ just after the
decay can be estimated by the energy density of the superheavy field, $\rho_X=m_X n_X$,
where $m_X$ and $n_X$ are the mass of $X$ and the number density of $X$, respectively.  
The reheating temperature $T_{RH}$ can be defined by the relation 
$\rho_{\text{rad}}=g_*(T_{RH})T_{RH}^4$ at the decay. 
The ratio of the dark matter number density, $n_{\chi}$, to the entropy density, $s$, produced 
by this direct decay process is estimated as
\begin{eqnarray}
\frac{n_{\direct}}{s}
&= \nu \frac{n_X}{s} = \nu \frac{1}{m_X} \frac{\rho_{\text{rad}}}{s}
= \frac{3}{4} \nu \frac{g_*(T_{RH})}{g_{*s}(T_{RH})} \frac{T_{RH}}{m_X} \\
&\approx \frac{3}{4}\nu \frac{T_{RH}}{m_X}
= 7.5 \times 10^{-12}\nu \left(\frac{T_{RH}}{1 \mev}\right)\left(\frac{10^{8} \gev}{m_X}\right),
\end{eqnarray}
where  $g_*$ and $g_{*s}$ are the number of freedoms of thermalized particles
for the radiation energy density and for the entropy density, respectively.
Usually, the number of produced LSPs per one $X$ decay, $\nu$, is taken as
$\nu\sim 1$.
In the ratio of the LSP density to the critical density,
\begin{equation}
\Omega_{\direct} h^2 
= 0.2 \nu\left(\frac{m_\chi}{100 \gev}\right)\left(\frac{T_{RH}}{1 \mev}\right) \left(\frac{10^{8} \gev}{m_X}\right),
\label{002}
\end{equation}
where $m_\chi$ is the mass of the LSP.
For the non-thermal production, the reheating temperature $T_{RH}$ must be
larger than O(1 MeV) not to spoil the success of the Big Bang Nucleosynthesys (BBN)
and must be smaller than the
freeze-out temperature of the WIMP which is roughly given by the relation
$T_{freeze-out}\sim m_\chi/20$. 
If the reheating temperature is 5 GeV, then the superheavy mass must be taken as
$10^{12}$ GeV in order to obtain the observed dark matter abundance.

\section{Effect of shower}
If the mass of the superheavy particle is much larger than the SUSY breaking scale, then
even the showers produced by the primary decay modes include additional SUSY particles. If $\nu$ becomes much larger than
the naive expectation, $1$, then the above estimation for the dark matter abundance must 
be changed. In this section, we estimate the parameter $\nu$ by using the generalized
DGLAP equations\cite{SHDecay}.

Let us briefly outline the physics on this calculation.
The primary decay products are the MSSM particles and 
generically off-shell. Since each MSSM particle in the primary decay modes has very large virtualities of order $m_X$, it produces a shower. In the shower, one virtual particle
 splits into two other particles with smaller virtualities.
When the virtuality is larger than 
 the SUSY breaking scale, SUSY particles are also produced in the shower.

  What we would like to know is how many SUSY particles appear in the shower produced by
the primary MSSM particle $I$ with $O(m_X)$ virtuality. Here, we call the number 
$\nu_I(m_X)$, since the number is dependent on the primary MSSM particle $I$.
Then the averaged $\nu$ is written as
\begin{equation}
\nu = \sum_f \text{Br} (X \to f) \sum_{I \in f} \nu_I (m_X),
\end{equation}
where $\text{Br} (X \to f)$ is the branching ratio of $X$ to the decay mode $f$. 
Once we know the numbers $\nu_I$ for all $I=$MSSM particles, we can apply the above 
relation for the averaged $\nu$ for any models in which the superheavy particle $X$
decays to the MSSM particles. We calculated these parameters basically following the
technique developed in Refs. \cite{SHDecay}. The parameters we calculated are listed
in Table 1 and 2.
\begin{table}[h]
\begin{center}
\begin{tabular}{|c|c|c|c|c|c|c|c|}
\hline
$Q [\text{GeV}]$ & $10^4$ & $10^5$ & $10^6$ & $10^7$ & $10^8$ & $10^9$ & $10^{10}$ \\
\hline
$q_L$ & $0.051$ & $0.16$ & $0.43$ & $1.1$ & $2.6$ & $6.1$ & $14$ \\
\hline
$\tilde{q}_L$ & $1.0$ & $1.1$ & $1.3$ & $1.9$ & $3.5$ & $6.9$ & $15$ \\
\hline
$q_R$ & $0.044$ & $0.14$ & $0.39$ & $0.97$ & $2.5$ & $5.8$ & $13$ \\
\hline
$\tilde{q}_R$ & $1.0$ & $1.1$ & $1.3$ & $1.8$ & $3.3$ & $6.6$ & $14$ \\
\hline
$l_L$ & $0.0087$ & $0.023$ & $0.048$ & $0.095$ & $0.19$ & $0.37$ & $0.74$ \\
\hline
$\tilde{l}_L$ & $1.0$ & $1.0$ & $1.0$ & $1.1$ & $1.2$ & $1.3$ & $1.7$ \\
\hline
$l_R$ & $0.0032$ & $0.0076$ & $0.014$ & $0.022$ & $0.036$ & $0.060$ & $0.10$ \\
\hline
$\tilde{l}_R$ & $1.0$ & $1.0$ & $1.0$ & $1.0$ & $1.0$ & $1.1$ & $1.1$ \\
\hline
$g$ & $0.12$ & $0.34$ & $0.88$ & $2.1$ & $5.2$ & $12$ & $28$ \\
\hline
$\tilde{g}$ & $1.0$ & $1.1$ & $1.6$ & $2.8$ & $5.8$ & $13$ & $28$ \\
\hline
$W$ & $0.041$ & $0.099$ & $0.19$ & $0.36$ & $0.72$ & $1.4$ & $3.0$ \\
\hline
$\tilde{W}$ & $1.0$ & $1.0$ & $1.1$ & $1.2$ & $1.5$ & $2.2$ & $3.7$ \\
\hline
$B$ & $0.013$ & $0.029$ & $0.049$ & $0.081$ & $0.15$ & $0.29$ & $0.64$ \\
\hline
$\tilde{B}$ & $1.0$ & $1.0$ & $1.0$ & $1.0$ & $1.1$ & $1.2$ & $1.6$ \\
\hline
$H_u$ & $0.017$ & $0.043$ & $0.082$ & $0.15$ & $0.29$ & $0.54$ & $1.1$ \\
\hline
$\tilde{H}_u$ & $1.0$ & $1.0$ & $1.0$ & $1.1$ & $1.2$ & $1.5$ & $2.0$ \\
\hline
$H_d$ & $0.017$ & $0.040$ & $0.074$ & $0.13$ & $0.23$ & $0.42$ & $0.80$ \\
\hline
$\tilde{H}_d$ & $1.0$ & $1.0$ & $1.0$ & $1.1$ & $1.2$ & $1.3$ & $1.7$ \\
\hline
\end{tabular}
\caption{The number of the LSP, $\nu_I$, in the parton shower by the MSSM particle $I$ 
with virtuality $Q$ ($m_X\sim 2 Q$). 
We take $\tan\beta\equiv \langle H_u\rangle/\langle H_d\rangle=3.6$. }
\end{center}
\end{table}

\begin{table}[h]
\begin{center}
\begin{tabular}{|c|c|c|c|c|c|c|}
\hline
$Q [\text{GeV}]$ & $10^{11}$ & $10^{12}$ & $10^{13}$ & $10^{14}$ & $10^{15}$ & $10^{16}$ \\
\hline
$q_L$ & $33(34)$ & $79(86)$ & $190(220)$ & $480(590)$ & $1200(1600)$ & $3200(4600)$ \\
\hline
$\tilde{q}_L$ & $34(35)$ & $80(86)$ & $190(220)$ & $480(590)$ & $1200(1600)$ & $3200(4500)$ \\
\hline
$q_R$ & $32(33)$ & $76(83)$ & $190(220)$ & $470(580)$ & $1200(1600)$ & $3100(4500)$ \\
\hline
$\tilde{q}_R$ & $33(34)$ & $77(84)$ & $190(220)$ & $470(580)$ & $1200(1600)$ & $3100(4500)$ \\
\hline
$l_L$ & $1.5(1.6)$ & $3.4(3.6)$ & $7.9(9.2)$ & $20(26)$ & $52(78)$ & $150(250)$ \\
\hline
$\tilde{l}_L$ & $2.5(2.5)$ & $4.3(4.6)$ & $8.8(10)$ & $21(26)$ & $53(78)$ & $150(250)$ \\
\hline
$l_R$ & $0.20(0.20)$ & $0.43(0.46)$ & $1.1(1.2)$ & $2.8(3.5)$ & $8.3(11)$ & $26(40)$ \\
\hline
$\tilde{l}_R$ & $1.2(1.2)$ & $1.4(1.4)$ & $2.0(2.2)$ & $3.8(4.5)$ & $9.3(12)$ & $27(41)$ \\
\hline
$g$ & $66(68)$ & $160(170)$ & $390(450)$ & $980(1200)$ & $2500(3400)$ & $6600(9800)$ \\
\hline
$\tilde{g}$ & $66(68)$ & $160(170)$ & $390(450)$ & $980(1200)$ & $2500(3400)$ & $6600(9600)$ \\
\hline
$W$ & $6.6(6.7)$ & $15(16)$ & $37(40)$ & $92(110)$ & $240(310)$ & $650(920)$ \\
\hline
$\tilde{W}$ & $7.3(7.4)$ & $16(17)$ & $37(41)$ & $93(110)$ & $240(310)$ & $650(920)$ \\
\hline
$B$ & $1.5(1.5)$ & $3.9(4.0)$ & $10(11)$ & $27(30)$ & $75(88)$ & $210(260)$ \\
\hline
$\tilde{B}$ & $2.4(2.4)$ & $4.8(4.9)$ & $11(12)$ & $28(31)$ & $76(90)$ & $210(270)$ \\
\hline
$H_u$ & $2.2(2.2)$ & $4.7(4.9)$ & $11(12)$ & $26(30)$ & $65(82)$ & $170(240)$ \\
\hline
$\tilde{H}_u$ & $3.1(3.1)$ & $5.6(5.7)$ & $12(12)$ & $27(30)$ & $66(80)$ & $170(230)$ \\
\hline
$H_d$ & $1.6(1.6)$ & $3.5(3.7)$ & $8.0(9.2)$ & $20(26)$ & $52(77) $ & $150(250)$ \\
\hline
$\tilde{H}_d$ & $2.5(2.5)$ & $4.3(4.6)$ & $8.9(10)$ & $21(26)$ & $53(78)$ & $150(250)$ \\
\hline
\end{tabular}
\caption{The number of the LSP, $\nu_I$, in the parton shower by the MSSM particle $I$ with
virtuality $Q$ ($m_X\sim 2 Q$). These are estimated by extraporating the 30(45) data points 
with $10^{-7}\leq x\leq 10^{-3}(0.1)$, which are calculated by generalized DGLAP
equations. We take $\tan\beta=3.6$. }

\end{center}
\end{table}

Let us explain briefly how to calculate these parameters. 
In order to calculate $\nu$, we introduce the fragmentation functions (FFs) $D_I^J(x, Q^2)$
$(0\leq x \leq 1, m_J^2\leq Q^2)$, where $D_I^J(x, Q^2)$ is the number density of particle $J$
with the energy $xQ$
which are produced in the parton shower by the primary field $I$ with initial virtuality 
$Q$. Then, the $\nu_I$ is estimated by
\begin{equation}
\nu_I(m_X) = \int_0^1 d x \; D_I^{J=LSP} (x, m_X^2).
\end{equation}
The fragmentation functions can be obtained as a solution of the DGLAP equation
\begin{equation}
\frac{d}{d \log (Q^2)} D_I^J (x,Q^2)
= \sum_{K \in \text{MSSM}} \frac{\alpha_{KI}(Q^2)}{2\pi} \int_x^1 \frac{d y}{y} \; P_{K \leftarrow I}(y) D_K^J (x/y,Q^2),
\end{equation}
where $P_{K \leftarrow I}(x)$ and $\alpha_{KI}(Q^2)$ are splitting functions (SFs)\cite{SF} on the 
three points interaction which produces $K$ in the MSSM and the running coupling constant 
for the interaction, respectively. As the boundary condition, 
\begin{equation}
D_I^J (x,m_J^2) = \delta^J_I \delta(1-x)
 \end{equation}
is imposed. This condition is to require that on-shell particle does not produce new
particle.
In order to solve these DGLAP equations, the generalized FFs 
$\tilde{D}_I^J(x,Q^2;Q_0^2) (m_J \leq Q_0 \leq Q)$ are introduced, which are defined as
solutions of DGLAP equations with the boundary condition
$\tilde{D}_I^J(x,Q_0^2;Q_0^2) = \delta_I^J \delta(1-x)$.
The usual FFs can be written as
$D_I^J(x,Q^2) = \tilde{D}_I^J(x,Q^2;m_J^2)$.
It is quite useful that the usual FFs can be decomposed as
\begin{equation}
D_I^J (x,Q^2)
= \sum_{K \in \text{MSSM}} \int_x^1 \frac{d y}{y} \; \tilde{D}_I^{K} (y, Q^2; Q_0^2) D_{K}^J (x/y, Q_0^2),
\end{equation}
if $Q_0$ is taken as the SUSY breaking scale, the electroweak symmetry breaking scale, or
hadronization scale, etc. For example, if we take $Q_0$ as the SUSY breaking scale $m_{SUSY}$,
then, the virtuality is always larger than the SUSY breaking scale in the DGLAP equations, 
$Q\geq m_{SUSY}$, and therefore, it is sufficient to solve the SUSY DGLAP equation to obtain
$\tilde{D}_I^J(x,Q^2;m_{\text{SUSY}}^2)$. In other words, we can neglect the SUSY breaking
effect in calculating $\tilde{D}_I^J(x,Q^2;m_{\text{SUSY}}^2)$. Since the above equation can
be rewritten as
\begin{equation}
\int_0^1dx D_I^J (x,Q^2)
= \sum_{K \in \text{MSSM}} \int_0^1d y \; \tilde{D}_I^{K} (y, Q^2; Q_0^2)
\int_0^1 d z D_{K}^J (z, Q_0^2),
\end{equation}
the parameters $\nu_I$ can be obtained as
\begin{eqnarray}
\nu_I(m_X)&=&\int_0^1dx D_I^{J=LSP} (x,m_X^2)
= \sum_{K \in \text{MSSM}} \int_0^1d y \; \tilde{D}_I^{K} (y, m_X^2; m_{\text{SUSY}}^2)
\int_0^1 d z D_{K}^J (z, m_{\text{SUSY}}^2)  \nonumber \\ 
&= & \sum_{K \in \text{sparticles}} \int_0^1d y \; \tilde{D}_I^{K} (y, m_X^2; m_{\text{SUSY}}^2). 
\label{nu}
\end{eqnarray}
Here, in the last equality, we use the assumption that one sparticle with virtuality
$Q=m_{\text{SUSY}}$ produces only one LSP in the decay, i.e., 
$\int_0^1 d z D_{K}^J (z, m_{\text{SUSY}}^2)=1$ for any sparticles and the SM particle
with virtuality $Q=m_{\text{SUSY}}$ produces no LSP.

In order to estimate the generalized FFs, we use the program "SHDecay" in which
the generalized SUSY DGLAP equations are numerically solved\cite{SHDecay}. 
The program "SHDecay" includes the all gauge interactions in the SM and the third
generation Yukawa couplings. We calculated the parameters $\nu_I$ for the third generation
fields and the first two generation fields separately, but the calculated values are
the almost same, so in Table 1 and 2, we do not distinguish the third generation fields
from the first two generation fields.  
We have several remarks on the calculation.
First, the integration in eq. (\ref{nu}) has infra-red divergence, because we solved the
generalized DGLAP equations in SUSY limit. 
The number of sparticles with smaller
energy than their mass becomes quite large. Of course, this situation is unphysical.
We just introduce the infra-red cutoff for the parameter $y$ as 
$y_{min}=m_{\text{SUSY}}/m_X$. (We took $m_{\text{SUSY}}=1$ TeV.)
Second, in the "SHDecay", the FFs can be calculated until $y=10^{-7}$. This is because
of the reliablity of the perturbation. It has been noted in Ref.\cite{SHDecay} 
that the energy conservation can 
be checked in this calculation in the 1 percent level. Actually, main contribution
to the energy comes from the FFs with larger $y$, but for the number of the produced
particles the FFs with smaller $y$ is more important. For $m_X\leq 10^{10}$ GeV, we
can calculate the parameter $\nu$ by integrating the FFs directly (see Table 1), 
but for $m_X>10^{10}$ GeV, we have no data. Therefore, we just assume that the
$\tilde D_I^J=A_I^J y^{\alpha_I^J}$, where the $\alpha_I^J$ and $A_I^J$ are 
determined by fitting the 
FFs $\tilde D_I^J$ with 30(45) points between $10^{-7}\leq y\leq 10^{-3}(0.1)$, and 
we can obtain the parameters $\nu_I$
as 
\begin{equation}
\nu_I=\nu_I(10^{-7}\leq y \leq 1)+\sum_{K \in \text{sparticles}} \int_\epsilon^{10^{-7}}d y \; \tilde{D}_I^{K} (y, m_X^2; m_{\text{SUSY}}^2),
\end{equation}
where $\epsilon=m_{\text{SUSY}}/m_X$. The results are shown in Table 2.
Unfortunately, this approximation is not so good especially for large $Q$, as seen in Table 2
that the estimated values for the two different fitting regions have larger discrepancies
for larger $Q$.  
This is because the FFs in smaller $x$ are more important
for the estimation of the number of produced sparticles in the shower.

As seen in Table 1 and 2, there is a rough relation between $\nu_I$ and $\nu_{\tilde I}$ as
$\nu_{\tilde I}\sim \nu_I+1$, which is reasonable because the difference of the number of 
primary sparticle is just one.

It is interesting that the upper bound of the superheavy mass for obtaining the observed DM
abundance can be increased from $10^{12}$ GeV to $10^{14}$ GeV, because the inflaton mass
for the chaotic inflation is around $10^{13}$ GeV. For the superheavy field with the mass
around $10^{13}$ GeV, the ratio of the density can be written as
\begin{equation}
\Omega_{\direct} h^2 
= 0.2 \left(\frac{\nu}{100}\right)\left(\frac{m_\chi}{100 \gev}\right)\left(\frac{T_{RH}}{1 \gev}\right) \left(\frac{10^{13} \gev}{m_X}\right).
\label{13}
\end{equation}
If the reheating temperature of the inflation is around 1 GeV, non-thermal production of the LSP
can explain the observed DM abundance.

In the heavy gravitino scenario in which the cosmological moduli problem can be solved because
the lifetime of the moduli fields becomes shorter than 1 second, the $\nu$ values for $Q\sim 10^{4-5} \gev$ must be important. In Ref. \cite{NTDM3}, it is noted that 
in order to produce the observed abundance
of the Wino-like LSP, $\nu\sim 10^{-4}$ is required. However, in our calculation,
especially, for the moduli which decays to gluons (Higgs) in certain portion, 
$\nu$ can be O(0.1)(O(0.01)), which is too large
to obtain the observed DM abundance. 
\footnote{This value for $\nu\sim 0.1$ is reasonable, because there are
several processes to produce sparticles through virtual gluon, for example, 
$X\rightarrow gg^*\rightarrow g {\tilde q} {\bar{\tilde{q}}}$, where $g$ and $\tilde q$ are 
gluon and squark, respectively. This process is suppressed because of the three body decay,
but the number of the final squarks is large, so the branching ratio for the process can 
be large.}

\section{Discussion and summary}
In principle, also in the process of the thermalization of the high energy particles, 
sparticles can be produced. Actually, the hard collision between the high energy particle
and a particle in the thermal bath can produce the sparticles if the center of mass energy
is larger than the sum of the produced sparticle masses. 
If the energy of the high energy particle, $E$, is larger than $m_{SUSY}^2/T_{RH}$,
such processes can be expected.  
However, as discussed in Ref.\cite{NTDM5}, such production is negligible for the LSP non-thermal 
production,
because the thermalization of the high energy particles is so rapid through soft processes
that the hard collision rarely happens before the thermalization 
finishes\cite{thermalization1}-\cite{thermalization3}.

In summary, we calculate the averaged number of the LSPs produced by the decay of a single
superheavy field $X$ by using generalized DGLAP equations. The number $\nu$ can be O(100)
for colored primary particle with the virtuality $Q\sim O(10^{13} \gev)$. As the result, 
even by the non-thermal production through the decay of chaotic inflaton with the mass, 
$O(10^{13} \gev)$, the observed abundance
 of the DM can be obtained. Moreover, even if the primary decay modes of 
 the heavy moduli with the mass, $O(100 \tev)$,
include only some of the standard model particles like gluons or
Higgses, the LSPs can produce in the shower, which leads to $\nu\sim O(0.1)$ or $O(0.01)$,
which is larger than $10^{-4}$ which is required to obtain the observed abundance of the LSPs.
 
In huge parameter region, the DM produced through the decay of the superheavy field $X$
 is over produced if the energy density of the superheavy 
fields dominates the energy density of the universe.
However, if the produced DM abundance is larger than the thermal abundance of the DM, then the pair annihilation 
process can reduce the abundance. The thermal abundance of the DM after the pair annihilation
can be estimated as
\begin{equation}
\Omega_{LSP}h^2\sim 0.25\times \left(\frac{m_{LSP}}{100\gev}\right)^3
\left(\frac{10^{-3}}{m_{LSP}^2\langle \sigma v\rangle}\right)\left(\frac{100\mev}{T_{RH}}\right)\left(\frac{10}{g_*(T_{RH})}\right)^{\frac{1}{2}}
\end{equation}
by solving the Boltzumann equations\cite{annihilation}.
Since the relation $m_{LSP}^2\langle \sigma v\rangle\sim 10^{-3}$ is typical for the Wino LSP or Higgsino LSP,
then O(100 MeV) reheating temperature can realize the observed value for the DM abundance in that case.

The calculation of the DM abundance in the case in which the energy of the superheavy field does not dominate
the energy of the universe is straighforward. By multiplying the ratio $\rho_X/\rho_{rad}$ at the decay time to
the equations for the DM abundance, we can obtain the results.

We hope our calculation can be applied into many cases in which non-thermal production of
DM are taken into account. 

\section*{Acknowledgments}
We thank S. Matsumoto and J. Hisano for valuable comments.
N.M. is supported in part by Grants-in-Aid for Scientific Research from
MEXT of Japan.
This work was partially supported by the Grand-in-Aid for Nagoya
University Global COE Program,
``Quest for Fundamental Principles in the Universe:
from Particles to the Solar System and the Cosmos'',
from the MEXT of Japan.

\end{document}